\documentclass{article}

\PassOptionsToPackage{numbers, compress}{natbib}

 \usepackage[preprint]{neurips_2026}

\usepackage[utf8]{inputenc} 
\usepackage[T1]{fontenc}    
\usepackage{hyperref}       
\usepackage{url}            
\usepackage{booktabs}       
\usepackage{amsfonts}       
\usepackage{nicefrac}       
\usepackage{microtype}      
\usepackage{xcolor}         
\usepackage{rotating}
\usepackage{array}
\usepackage[most]{tcolorbox}

\usepackage{amsmath}
\usepackage{graphicx}
\usepackage{tabularx}
\usepackage{makecell}
\usepackage{array}
\usepackage{pifont}

\newcolumntype{L}[1]{>{\raggedright\arraybackslash}p{#1}}

\usepackage{xspace}
\newcommand{\benchnamelong}{\textsc{BlueFin}\xspace}
\newcommand{\benchname}{\textsc{BlueFin}\xspace}

\title{\benchnamelong: \\Benchmarking LLM Agents on Financial Spreadsheets}

\author{
  Srivatsa Kundurthy\textsuperscript{*\,1,2} \And
  Clara Na\textsuperscript{*\,3} \And
  Colton Moraine\textsuperscript{1} \And
  Anoushka Mohta\textsuperscript{1} \And
  Case Winter\textsuperscript{1} \And
  George Fang\textsuperscript{1} \And
  John Ling\textsuperscript{1} \And
  Emma Strubell\textsuperscript{3} \And
  Zach Kirshner\textsuperscript{1} \\[0.7em]
  \textsuperscript{1}Longitude Labs Inc.\qquad \textsuperscript{2}Cornell University\qquad \textsuperscript{3}Carnegie Mellon University
}

\begin{document}

\renewcommand{\thefootnote}{\fnsymbol{footnote}}
\maketitle
\footnotetext[1]{Equal contribution. Correspondence to \texttt{srivatsa@meridian.ai}.}
\renewcommand{\thefootnote}{\arabic{footnote}}
\setcounter{footnote}{0}

\begin{abstract}

We present \benchnamelong, a benchmark that tasks large language model (LLM) agents with synthesis, manipulation, and 
comprehension tasks over spreadsheet workbooks in the professional finance domain. 
Though estimates of the global population of paying users of spreadsheet software range in the hundreds of millions -- an order of magnitude more than the estimated global population of professional developers -- comparatively fewer resources have been devoted to exploring and expanding LLM capabilities in the spreadsheet domain, with fewer still dedicated to mirroring real occupational tasks encountered by those in professional finance roles. 
In response, we curate a set of 131 challenging, complex tasks with real-world relevance in the domain, containing 3,225 granular rubric criteria; 
notably, our rubric criteria and LM judge evaluations are validated by a team of expert human annotators, resulting in high-quality, granular evaluations of complex tasks that are difficult to verify programmatically but can be reliably evaluated by an LM judge agent. Our judge achieves parity with expert consensus ($\alpha=0.826$) with a macro-F1 score of 0.839. 
Frontier LLMs demonstrate poor performance on the challenging benchmark, with the strongest LLMs achieving less than 50\% average scores across tasks -- models exhibit particular weaknesses in dynamic correctness. 
Our contributions include a dataset of examples across three categories of spreadsheet tasks, an open source harness and agentic evaluation framework, and a characterization of existing frontier models' performance on our benchmark.
\end{abstract}

\section{Introduction}
\label{sec:introduction}

\begin{figure}[h]
\centering
\includegraphics[width=0.9\textwidth]{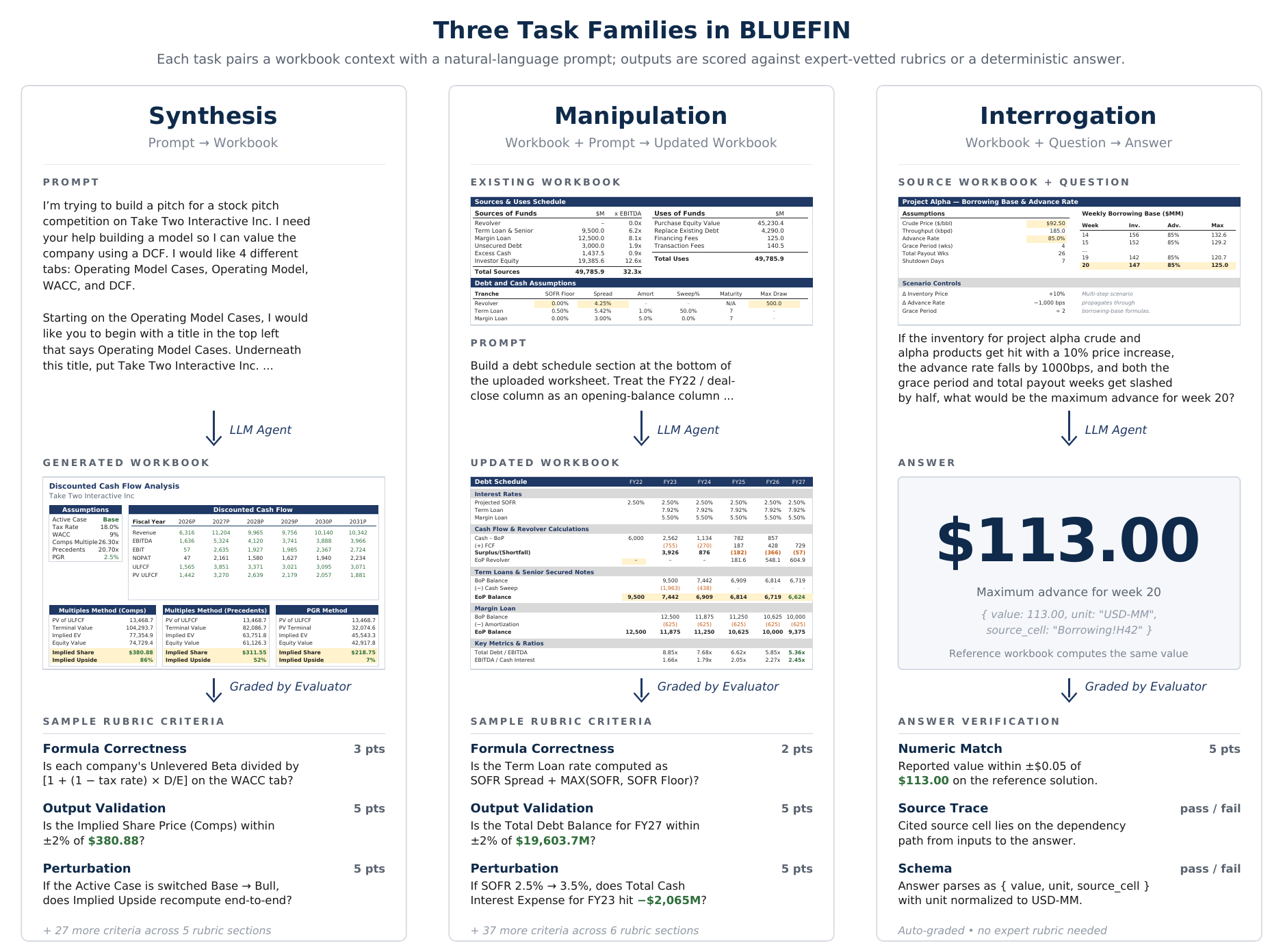}
\caption{BlueFin is a challenging benchmark for characterizing LLM spreadsheet generation.} 
\label{fig:fig1}
\end{figure}

Professional finance is among the most spreadsheet-intensive areas of the knowledge economy. According to the United States Bureau of Labor Statistics \citep{bls2024oes}, there were nearly 2.5 million professional finance workers in the US in 2024 who performed labor representative of hundreds of billions of dollars in compensation,\footnote{We arrive at our estimate of 2.5 million finance professionals by considering Accountants and auditors (1,448,290), Property appraisers and assessors (59,070), Budget analysts (47,170), Credit analysts (67,370), Financial and investment analysts (340,580), Personal financial advisors (270,480), Financial risk specialists (56,320), Financial examiners (62,830), and Financial specialists, all other (127,450).} with spreadsheets featuring heavily in their work; industry reports estimate that nearly all FP\&A professionals use spreadsheets weekly,\footnote{\href{https://www.financialprofessionals.org/training-resources/resources/articles/Details/useful-excel-functions-for-financial-planning-and-analysis}{https://www.financialprofessionals.org/training-resources/resources/articles/Details/useful-excel-functions-for-financial-planning-and-analysis}} and skilled financial analysts spend 10 hours a week on spreadsheet work.\footnote{\href{https://www.datarails.com/wp-content/uploads/2022/09/2022-CFO-Budgeting-and-Forecasting-Survey-Report.pdf}{https://www.datarails.com/wp-content/uploads/2022/09/2022-CFO-Budgeting-and-Forecasting-Survey-Report.pdf}} Previous work has shown that professional spreadsheet work is error-prone, and errors have the potential for outsized consequences;  \citet{powell2008impacterrorsoperationalspreadsheets} found 117 errors across an audit of 50 operational spreadsheets, constituting 0.8\% to 1.8\% of all formula cells and leading to error impacts of up to \$100,000,000. Other documented incidents include a spreadsheet formula error contributing to JPMorgan's \$6B+ ``London Whale'' loss, and spreadsheet errors leading to Fannie Mae's 2003 \$1.1B restatement.

Recent developments in spreadsheet capabilities by frontier labs (e.g. Claude for Excel,\footnote{\url{https://claude.com/claude-for-excel}} ChatGPT for Excel\footnote{\href{https://chatgpt.com/apps/spreadsheets/}{https://chatgpt.com/apps/spreadsheets/}}) suggest ongoing investment in the space. 
Previous studies report that generative AI use for work tasks can facilitate greater quality and speed \citep{noy2023experimental, dellacqua2026jagged} in work tasks, with the greatest productivity benefits conferred to less experienced professionals. On the other hand, \citet{dellacqua2026jagged} note that AI use impedes productivity when the task's difficulty exceeds the AI tools' capabilities. Thus, it is imperative that we have adequate infrastructure for evaluating AI system capabilities in this domain.

However, to our knowledge, no existing benchmark is well-suited for reliable evaluation of multi-step spreadsheet tasks matching the multi-dimensional challenges of real-world professional finance tasks. The relatively sparse landscape of evaluations in this domain (compared to e.g. coding benchmarks for complex software engineering tasks \citep{jiminez2024swebench, chowdhury2024swebenchverified, deng2025swe}) reflects multiple inherent challenges. (1) Enterprise financial spreadsheets often contain corporate secrets or sensitive information on clients, complicating procurement of documents for an appropriate dataset. Moreover, (2) it is nontrivial to evaluate successful task completion in a setting where the output must not only be structurally and numerically correct but also satisfy user-centric usability and clarity criteria that are not programmatically verifiable. 

Prior benchmarks and evaluations of spreadsheet tasks tend to evaluate relatively shallow capabilities such as formula generation \citep{zhao2024nl2formula} or atomic operations on pre-existing workbooks \citep{li_sheetcopilot_2023, zhang-etal-2025-tablellm, chen2025SheetRMagent}, assume largely flat, CSV-like tables \citep{zhao2024nl2formula, zhang-etal-2025-tablellm, chen2025SheetRMagent}, or limit scope to question-answering over fixed spreadsheets \citep{ravnik2026finsheetbenchsimplelookupscomplex} that are programmatically verifiable, and/or compress evaluation of an inherently complex task into a digestible metric or score vector. 
\citet{kundurthy2026spreadsheetarena} report that human preference evaluations tend to reflect obvious deficits in utility or presentation, but do not reliably capture domain specific conventions for finance spreadsheets. Likewise, existing benchmarks of skills in the finance domain focus on entirely different types of finance tasks (e.g. not involving spreadsheets), include only a handful of relevant examples as part of a broader effort to benchmark multiple occupational capabilities, limit tasks to those that are programmatically verifiable, and/or feature relatively simplified rubric criteria \citep{wang2024officebenchbenchmarkinglanguageagents, patwardhan2025gdpvalevaluatingaimodel, luo2026alphabench, opsahlong2026officeqaproenterprisebenchmark, bankertoolbench2026, vidgen2026apexagents, ravnik2026finsheetbenchsimplelookupscomplex}.

We address this gap by introducing \benchnamelong, a benchmark that tasks LLM agents with synthesis, comprehension, and manipulation tasks over spreadsheet workbooks in the professional finance domain.\footnote{The name \benchnamelong is loosely inspired by the idea of \textbf{b}ui\textbf{l}ding, \textbf{u}nderstanding, and \textbf{e}diting \textbf{fin}ancial spreadsheets.} \benchname is distinguished by: (1) \textit{high-quality, pragmatic tasks} that reflect the day-to-day responsibilities of finance professionals, e.g. building financial models from scratch, modifying existing models, and examining artifacts to inform decision-making, with a proportional emphasis on modification tasks; (2) focus on \textit{spreadsheets as salient artifacts} across all tasks; (3) depth of coverage in a specialized domain requiring \textit{complex, multi-step workflows}; and (4) \textit{fine-grained, granular evaluation} criteria vetted by domain experts. Creation of \benchname entailed hundreds of domain expert hours across task creation, annotation, refinement, and review, despite AI assistance with rubric creation. \benchname contains 3,225 granular rubric criteria over 10 end-to-end \textbf{synthesis} tasks (spreadsheet creation given natural language prompts), 82 patch-based \textbf{manipulation} tasks (spreadsheet modification according to a natural language prompt, given an input spreadsheet), and 39 \textbf{interrogation} tasks (comprehension and question answering tasks given an input spreadsheet). Of these, we fully release 305 rubric criteria over 11 tasks with a proportional split across task types. Manipulation tasks are most heavily featured in the benchmark, mirroring the responsibilities of investment banking (IB) and private equity (PE) analysts -- manipulation tasks in \benchname require at least 45+ minutes of work for an analyst to complete from scratch.

We characterize existing frontier models' performance on our benchmark, finding that frontier models achieve below-50\% average scores across diverse expert rubric criteria, with particular challenges in output validity and robust formula correctness (via perturbation). We open source a test harness and agentic evaluation framework for \benchname (\S\ref{sec:methodology}). Our hope is that \benchname inspires further interest from the community in complex spreadsheet tasks with specialized domain relevance, and that our detailed methodological descriptions can inform future work on benchmarks in highly specialized domains that have traditionally been challenging to build high quality benchmarks for.

\section{Related Work}
\label{sec:related_work}

There is a large and growing interest in exploring and developing AI systems' capabilities in specialized domains where satisfactory task completion must be grounded in structured artifacts and domain-specific conventions. In the finance domain, where \benchnamelong is situated, spreadsheet workbooks are a prominent structured artifact that relevant tasks frequently interface with; domain-specific conventions \citep{FAST2015} influence expected best practices around finance spreadsheets. 

Previous efforts and investment have focused heavily on the code domain; numerous benchmarks evaluate code generation and program synthesis capabilities in settings where candidate outputs are largely programmatically verifiable via code execution \citep{chen2021evaluatinglargelanguagemodelscodex}, even for increasingly complex tasks typically approached with agentic systems \citep{jiminez2024swebench, chowdhury2024swebenchverified, deng2025swe}. In general, code artifacts are well-represented in training corpora \citep{kocetkov2022stack3tbpermissively, wang-etal-2023-mconala} such that frontier models are often specifically trained to perform well in the setting \citep{chen2021evaluatinglargelanguagemodelscodex, Li_2022_alphacode, roziere2024codellamaopenfoundation}.

In the spreadsheet and professional finance domains, the existence and increasing complexity of tasks included in recent AI benchmarks reflect an ongoing investment in the space and/or general interest in the spreadsheet as a structured artifact that constitutes a compelling setting for evaluating AI systems' reasoning and instruction following capabilities. Like code, spreadsheets can have dense, graph-structured dependencies, and errors can be syntactic or non-obvious \citep{panko2010taxonomy}. However, professional finance tasks over spreadsheets present challenges that are distinct from code domains and arguably warrant further study: complex spreadsheet tasks require 2D spatial reasoning across cells that is less inherent to code settings. Visual stylistic and formatting concerns not applicable to code are featured heavily in spreadsheets (and are often salient factors in human preference evaluations of generated spreadsheets \citep{kundurthy2026spreadsheetarena}), and standard financial modeling conventions \citep{FAST2015} are less relevant to ``general'' spreadsheet tasks that more existing benchmarks have engaged with. Evaluation presents a challenge for complex professional finance tasks; evaluation criteria are nontrivial to express as programmatically verifiable tests that are often possible with complex, repository-level coding tasks. Additionally, professional finance spreadsheets are difficult to procure high-quality, realistic examples for, due to licensing and privacy concerns that are less prevalent in the software domain due to widespread open-sourcing and permissive licensing.

\benchnamelong addresses a gap in the literature, constituting a high-quality collection of complex tasks that call for multi-step reasoning and manipulation over realistic financial workbooks. Many existing spreadsheet benchmarks are limited to relatively simpler tasks that naturally lend themselves to programmatic verification such as spreadsheet formula generation \citep{zhao2024nl2formula} or atomic manipulation (e.g. ``Please highlight Sales between 200 and 500'' \citep{li_sheetcopilot_2023}) from detailed natural language instructions \citep{li_sheetcopilot_2023, ma2024spreadsheetbench, li2025mimotable, chen2025SheetRMagent} that tend to map cleanly to evaluation criteria . For instance, even SheetRM \citep{chen2025SheetRMagent}, which explicitly calls for long-horizon trajectories, modulates task difficulty by simply concatenating multiple relatively well-defined spreadsheet manipulation instructions. In contrast, a realistic spreadsheet benchmark in the professional finance domain should evaluate execution of complex end-to-end workflows that are expressed naturally (i.e. underspecified compared to existing spreadsheet benchmarks); a performant AI agent should be able to infer a viable plan and leverage appropriate tools to complete the task, satisfying both explicit specifications and implicit expectations relevant to the domain. Evaluation of such complex workflows is nontrivial as a result.

More recent works have explored AI system capabilities on similarly complex end-to-end workflows in specialized domains. Similarly to \benchname but in a different task setting that also does not easily lend itself to programmatic verification due to its complexity and subjectivity, \citet{sharma2025researchrubricsbenchmarkpromptsrubrics} curate a set of thousands of high quality, granular evaluation criteria on a relatively smaller set of Deep Research task instances. Multiple benchmarks such as GDPVal \citep{patwardhan2025gdpvalevaluatingaimodel}, OfficeBench \citep{wang2024officebenchbenchmarkinglanguageagents}, OfficeQA Pro \citep{opsahlong2026officeqaproenterprisebenchmark}, and APEX-Agents \citep{vidgen2026apexagents} also include a limited set of workflows relevant to finance professionals as parts of broader studies of AI capabilities on real-world end-to-end workflows -- in contrast, \benchname features a focus on complex professional finance tasks over spreadsheet artifacts. 

\citet{kundurthy2026spreadsheetarena} focus on end-to-end spreadsheet synthesis, noting that models often fail to produce satisfactory spreadsheets in the finance domain in particular; \benchname includes spreadsheet synthesis tasks but additionally contains rich examples across manipulation and interrogation tasks, which more fully reflects the spreadsheet workflows of a finance professional. 
Fin-SheetBench \citep{ravnik2026finsheetbenchsimplelookupscomplex} is a benchmark of 133 question answering tasks over 100 years of U.S. Treasury Bulletins, using synthetic spreadsheets. The benchmark includes a range of question difficulty levels, and the authors report degraded accuracy performance on larger, more complex files -- notably, our closed frontier models (which outperform the open weight models we benchmark) achieve only 34.7\%-50.0\% accuracy performance on interrogation tasks, such that our upper end of accuracy performance just barely reaches the \textit{lower} end of accuracy performance on \citet{ravnik2026finsheetbenchsimplelookupscomplex}'s tasks. We attribute the discrepancy to the complexity of our interrogation tasks: many tasks require agents to update spreadsheet assumptions, recompute multi-level dependencies, and interpret the resulting state to answer correctly. A concurrent work, BankerToolBench \citep{bankertoolbench2026}, is closely related in motivation and scope. \citet{bankertoolbench2026} create 100 tasks with even more (100+) rubric criteria per task, focusing on end-to-end workflows that faithfully represent workflows of real junior investment bankers. We view our work as highly complementary: empirical results on the benchmarks broadly support each other in that aggregate rubric scores span similar ranges (clustering in the 40-60\% ranges), with Grok and open models lagging behind other frontier models, and performance varying across rubric item and task types. 
Whereas BankerToolBench includes tasks across the spectrum of investment banker work tasks, such as creating discounted cash flow (DCF) models (in scope for \benchname) and preparing pitch decks (outside \benchname's scope), \benchname prioritizes coverage of spreadsheet tasks in the domain over a similar number of task instances -- facilitating in depth analysis of specific model capabilities over focused evaluation criteria.

\section{\benchnamelong}
\label{sec:methodology} 

We describe our benchmark creation process, which was used to develop both the public set and hidden set of tasks. We release public tasks at \url{https://huggingface.co/datasets/Longitude-Labs/bluefin-release}.

\subsection{Dataset Construction}
\label{subsec:dataset_construction}

Tasks for \benchname were constructed by vetted contributors with domain expertise, who were compensated for their work at substantially above prevailing market rates for annotation work, reflecting the level of
domain expertise required. 
Contributors underwent screening and both asynchronous and synchronous training before being eligible to submit tasks. Tasks themselves were vetted for quality and complexity, with a floor of at least 45 minutes of estimated work for an investment banker or private equity analyst to complete from scratch, for generative tasks; extensive multi-tab synthesis tasks such as full debt builds and merger models were estimated to require multiple hours of analyst work. For manipulation tasks, the most prominent task category in our benchmark, annotators were provided a seed workbook to start with -- seed spreadsheets ended up with an average of 1.5 tasks each. Given seed workbooks, contributors were asked to submit modified workbooks along with natural language instructions that would elicit the changes made -- contributors were asked to provide instructions ``as [they] would brief a capable analyst.'' Candidate rubric criteria were generated by Claude Opus 4.6 for the task submitted, and reviewers were tasked with also validating and refining the proposed rubric items. 

Similarly, for synthesis tasks, annotators were provided with a seed workbook hand-created by finance experts and asked to conceptualize a partial or full build mapping to a real-world deliverable an analyst would be asked to construct from scratch, also providing a natural language prompt along with the task-complete workbook. Additionally, a portion of our synthesis tasks were sourced from pre-existing models collected from industry experts. Though the set of six rubric item categories are consistent between synthesis and manipulation tasks, rubrics for synthesis tasks tend to emphasize presentation more heavily than in other tasks categories where seed notebooks already contain many desired presentation elements.

Interrogation tasks specified a correct answer and were not associated with rubrics. For interrogation tasks, we sourced a wide range of high-quality models reflecting professional use cases and standards and analyzed the dynamic flow of each model: what the main outputs were, how the supporting pieces contributed to those outputs, and where the underlying data was being pulled from. Given an understanding of model structure, we identified inputs that affected downstream outputs through multi-step calculation chains (vs. simple direct links), and we prioritized these examples to develop more difficult questions in our set.

\begin{figure}[h]
\centering
\includegraphics[width=\textwidth]{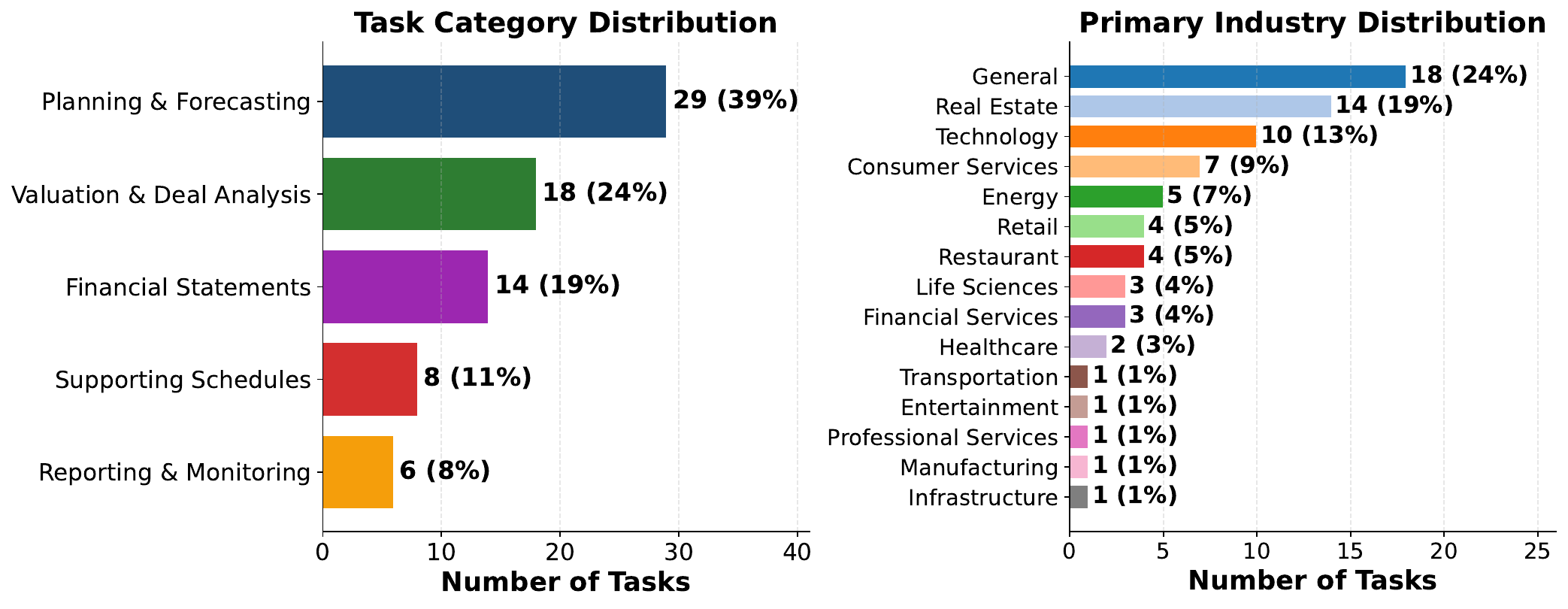}
\caption{Composition of the manipulation held-out set (n=75). Tasks span 5 primary financial-modeling categories (Planning \& Forecasting at 39\%, Valuation \& Deal Analysis at 24\%, Financial Statements at 19\%, Supporting Schedules at 11\%, Reporting \& Monitoring at 8\%) and 16 industries (Real Estate, Technology, Energy, Consumer Services, and Healthcare collectively 60\%; the remaining 11 industries each $\leq$5\%).} 
\label{fig:tasks_profile}
\end{figure}

Validated manipulation and synthesis tasks were tagged, also by Opus 4.6, with labels for \textit{primary industry} and \textit{task category}. Figure~\ref{fig:tasks_profile} displays the number of held-out manipulation tasks over task categories and industries. 

Table~\ref{tab:criteria-by-section} reports rubric criteria counts in 6 criterion categories across 10 synthesis (9 hold-out, 1 public) and 82 manipulation (75 hold-out, 7 public) tasks. See Appx.~\ref{app:bench_examples} for descriptions of rubric categories and examples of criteria over each of the categories.

\begin{table}[h]
\centering
\footnotesize
\caption{Rubric Criteria Counts by Section}
\label{tab:criteria-by-section}
\begin{tabular}{lrrrrr}
\toprule
Section             & Synth. (H) & Manip. (P) & Synth. (H) & Manip. (P) & Total \\
\midrule
Formula Correctness &       869  &       87  &      112   &      14   & 1{,}082 \\
Output Validation   &       628  &       65  &       62   &      10   &    765  \\
Perturbation        &       378  &       45  &       39   &       6   &    468  \\
Model Integration   &       335  &       30  &       14   &       4   &    383  \\
Presentation        &       281  &       25  &       26   &       2   &    334  \\
Pitfalls            &       150  &       14  &       26   &       3   &    193  \\
\midrule
Total               &    2{,}641 &      266  &      279   &      39   & 3{,}225 \\
\bottomrule
\end{tabular}
\end{table}

We estimate approximately 5 hours of annotator labor total per task. 
We obtained consent from dataset contributors to publicly release submitted task examples via full IP assignment. We retain a private set primarily out of potential leakage concerns. In addition to general quality checks on submitted tasks, we carefully check tasks for personally identifiable information (PII) and potential IP or licensing violations in submitted spreadsheets. See Appx.\ref{app:contributor_blurbs} for additional details about contributor instructions and annotation tasks.

\label{subsec:agent_judge}

\section{Evaluations}
\label{sec:evaluation}

We describe infrastructure and methodology for task evaluation, and report results of existing models on \benchname.

We evaluate five frontier models (\textbf{Claude Opus~4.7}, \textbf{Claude Sonnet~4.6}, \textbf{GPT-5.5}, \textbf{Gemini~3.1 Pro Preview}, and \textbf{Grok~4.20}) on the held-out portion of \benchname (120 tasks: 75 manipulation, 9 synthesis, 36 interrogation). All models are set to high reasoning via their respective provider APIs. See Appendix~\ref{app:public_set} for results on a publicly released set with an expanded set of models. Table~\ref{tab:cost_tokens} and Appendix~\ref{app:cost_tokens} discuss total costs associated with evaluation.

\textbf{Evaluation Harness.} We expose an agentic harness of 20 tools organized into six categories that cover the full lifecycle of spreadsheet construction. The functional surface offers fine-grained structured access: read primitives that return ranges and full workbook state, batch write primitives that mutate cells and create or delete sheets, formatting primitives that operate on either single ranges or batched format-spec arrays, and an explicit recalculation tool that drives a LibreOffice-headless calculation pass with iterative-calc enabled so debt schedules and LBO-style circular references converge correctly. Alongside these, a sandboxed code execution tool exposes the workbook as an in-memory openpyxl object with restricted built-ins and a tight import whitelist (no file I/O, no network); models optimized for programmatic manipulation or batch verification can perform bulk edits in a single call. Following the design philosophy of HELM~\cite{liang2023holisticevaluationlanguagemodels} and GAIA~\cite{mialon2023gaiabenchmarkgeneralai}, we adopt a minimal-elicitation prompt (61 words; see Appx.~\ref{app:si_harness}) with no workflow prescriptions, tool preferences, or domain conventions, isolating raw task-specific capability from instruction-following confounds~\cite{zhu2021tatqaquestionansweringbenchmark}. Details of harness implementation are included in Appendix~\ref{app:harness_methodology}. In general \benchname evaluates financial reasoning, tool use capabilities, and domain convention adherence, under a sufficiently expressive but relatively minimal harness, though we acknowledge that more complex scaffolding has the potential to extract stronger task performance.

\textbf{Agentic Grading.} The generative tasks (synthesis, manipulation) are graded by an agentic LLM judge that interacts with the output workbook through the same 20-tool spreadsheet environment used by the agent under evaluation. However, the System Instruction is strengthened to provide opinionated domain conventions and financial instructions that enable the judge to serve as a proxy expert human rater behavior (see appendix~\ref{app:si_harness}). Given a model's \texttt{output.xlsx} and the task rubric, the judge issues a sequence of tool calls and code execution commands to inspect both static structure (e.g. ``does cell B19 contain a SUM formula?'' ``does the DCF tab exist?'') and dynamic behavior (e.g. ``when input X is changed to Y, does output A update to B within tolerance?''), and emits a binary decision per criterion with mandatory short-form evidence pointing to the cells observed. Scores aggregate as the weighted ratio of points met to maximum points (0–100). For our LM judge, we use GPT-5.4 with reasoning\_effort=high under an expanded, domain-specific SI. GPT-5.4 was selected and validated via comparison to expert labels established via an inter-annotator alignment study over 384 binary criteria where we observed a Krippendorff's Alpha of 0.826. All three candidate judges aligned with expert consensus at comparable levels near the human IRR reliability ceiling, though GPT-5.4's macro F1 score of 0.839 was higher than Gemini 3.1 Pro's (0.803) and Sonnet 4.6's (0.805).

All code is released for reproduction of results (and surveying of configs and system prompts) at \href{https://github.com/Longitude-Labs/bluefin}{https://github.com/Longitude-Labs/bluefin}.

\subsection{Model Overall Performance}
\label{subsec:top_line}

\begin{figure}[h]
\centering
\includegraphics[width=0.9\textwidth]{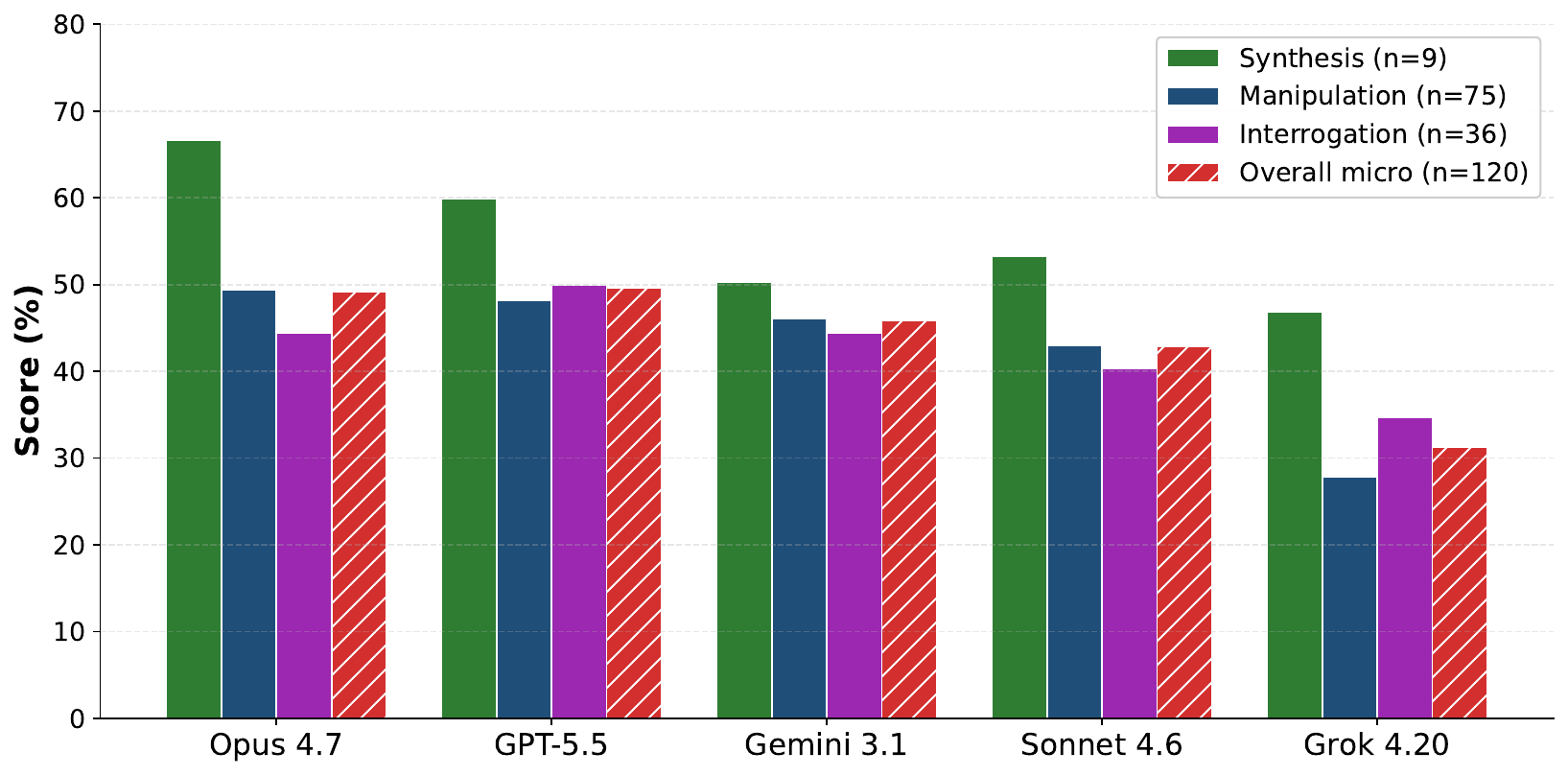}
\caption{Held-out performance per task type. Even the strongest frontier models remain below 50\% on the benchmark overall, despite the same models rapidly saturating such frontier capability benchmarks as  SWE-bench Pro \cite{deng2025swe}.}
\label{fig:topline}
\end{figure}

Table~\ref{fig:topline} reports the performance of five frontier models on the held-out set; Figure~\ref{fig:topline} visualizes the breakdown. 
See Table~\ref{tab:public_results} in Appx~\ref{app:public_set} for results on the public set on an expanded set of models, including 2 open-source candidates, Kimi K2.6 \cite{kimiteam2026kimik25visualagentic} and Minimax M2.5 \cite{minimax2026m25}. 
\begin{table}[h]
\centering
\footnotesize
\begin{tabular}{l|c|ccccc}
\toprule
\textbf{Set} & \textbf{N} & \textbf{Opus} & \textbf{Sonnet} & \textbf{GPT-5.5} & \textbf{Gemini} & \textbf{Grok} \\
\midrule
Manipulation  & 75  & 49.4 & 43.0 & 48.2 & 46.1 & 27.8 \\
Synthesis   & 9   & \textbf{66.7} & 53.2 & 59.9 & 50.3 & 46.9 \\
Interrogation & 36 & 44.4 & 40.3 & \textbf{50.0} & 44.4 & 34.7 \\
\midrule
\textbf{Overall} & 120 & 49.2 & 42.9 & \textbf{49.6} & 45.9 & 31.3 \\
\bottomrule
\end{tabular}
\caption{Held-out performance (\%, score is criteria-weighted percentage; non-completions count as zero). No frontier model exceeds 50\% performance. }
\label{tab:topline}
\end{table}

Notably, we find that frontier closed-source models all struggle to clear the 50\% threshold, with GPT-5.5 and Opus 4.7 at the forefront. We find this profile is distinct across different task types, as captured in Figure~\ref{fig:topline}. Models pass the most rubric items on Synthesis tasks, which we believe may reflect the relative flexibility of building a model from scratch compared with modifying an existing workbook whose formulas, layouts, and references impose brittle integration constraints. However, manipulation tasks capture the bulk of spreadsheet work in the professional finance domain, and even top models demonstrate poorer ability to reason and manipulate in a manner that is well-integrated within the starting workbook. Interrogation tasks were observed to be challenging for models to answer perfectly. 

Generally, rubric scores correspond to a proportion of critical evaluation criteria passed. 
In particular, we view \benchname scores as a weighted snapshot over dimensions IB and PE practitioners consider essential in real workflows -- logic, integration, dynamic behavior, and errors -- without prescribing a single canonical layout. Alongside our aggregated rubric scores, we additionally explore model-generated outputs' utility to finance professionals in Appx.~\ref{app:human_utility}. 

\subsection{Characterizing Failure Modes}

\begin{figure}[h]
\centering
\includegraphics[width=\textwidth]{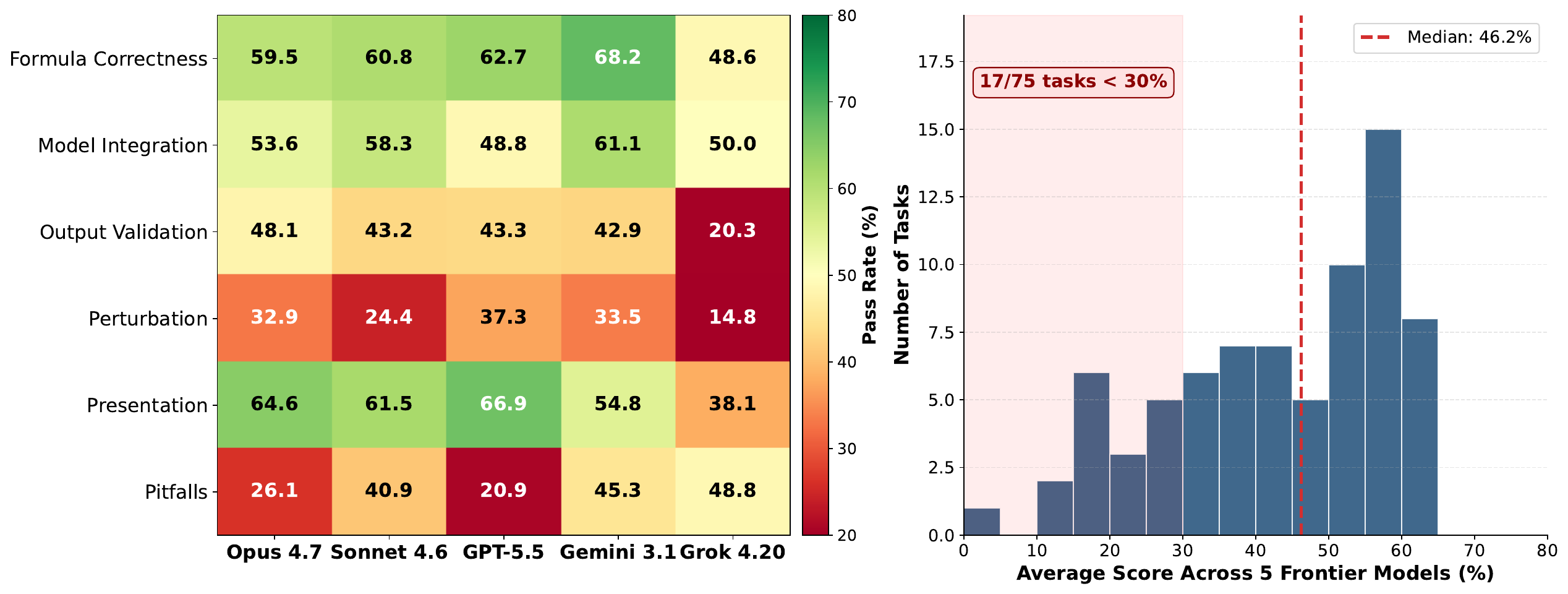}
\caption{(a) Per-section criterion pass rate across the five frontier models on the 75-task held-out set of manipulation tasks. Models are most reliable at structural, syntactically-checkable sections (Formula Correctness, Presentation) and weakest at sections that test computed numeric values (Output Validation) and dynamic behavior under perturbation. (b) Distribution of per-task average scores across the same 75 tasks (averaged across five frontier models). The median task sits at 46.2\%; only 44\% of tasks clear 50\% and not a single task clears 70\%. The 17/75 tasks (23\%) sitting below 30\% are genuinely difficult construction and modification problems where the typical frontier model fails more than two-thirds of rubric criteria.}
\label{fig:difficulty_combined}
\end{figure}

Decomposing rubric judgments by rubric section reveals a model-agnostic failure pattern, summarized in Figure~\ref{fig:difficulty_combined}(a). Out of the categories of rubric criteria, models most commonly pass \emph{Formula Correctness} criteria (50-68\% across the five frontier models). 
However, they pass fewer \emph{Output Validation} (20-48\%) and \emph{Perturbation} (15-37\%) criteria. In general, we observe a roughly 30 point gap between performance on formula correctness criteria, which check surface forms of static cell logic, vs dynamic-behavior correctness criteria, which evaluate robustness of model builds.

\textit{Perturbation} items expose weaknesses in dynamic behavior. A perturbation criterion specifies an input mutation (e.g., ``change WACC from 8.5\% to 10\%'') and an expected output shift (e.g., ``Enterprise Value moves from \$3{,}820M to within $\pm$2\% of \$4{,}205M''). The judge mutates the input, calls \texttt{recalc\_workbook}, and reads the target output. Models that hardcode intermediate values, erroneously overwrite a formula cell with a constant, or build partial circular structures that fail to converge, would also fail the perturbation check.

\textit{Output Validation} is an additional bottleneck. Even when formulas are structurally correct, the computed numeric values frequently miss the $\pm$2\% rubric tolerance. Manual inspection of low-scoring trajectories surfaces four recurring causes: sign errors on cash-outflow line items (adding where subtraction is required), date-axis misalignment (Q1-2024 vs.\ Q1-2025 reporting offsets), rate-vs-amount confusion (a 5\% spread encoded as 5.0 instead of 0.05), and end-of-period vs.\ beginning-of-period discounting in DCFs. 

\emph{Pitfalls} is a penalty section, where a higher pass rate means fewer detected errors. Grok's high pitfall rate reflects its tendency to produce less content overall, resulting in fewer potential triggers for pitfall criteria.

Figure~\ref{fig:difficulty_combined}(b) presents aggregate scores on the distribution of 5-model-average task scores across the 75-task held-out manipulation set. The median task averages \textbf{46.2\%}; \textbf{only 33 of 75 tasks (44\%) have a mean score of at least 50\%}, and \textbf{all task scores are below 70\%}. At the left tail, \textbf{17 of 75 tasks (23\%) sit below 30\%}.

\subsection{Cost-Performance Tradeoffs}
\label{subsec:cost_pareto}

\begin{figure}[h]
\centering
\includegraphics[width=0.75\textwidth]{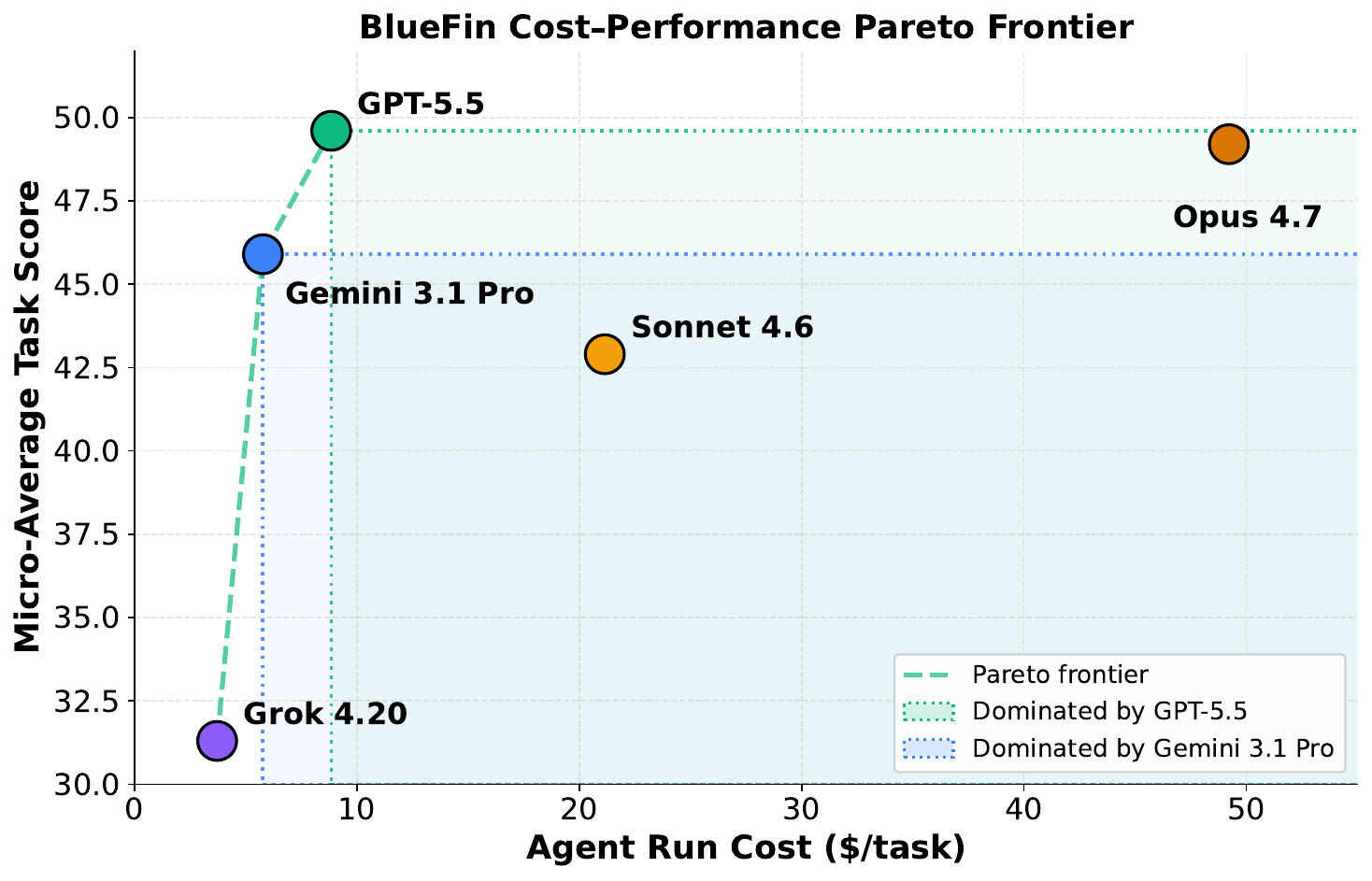}
\caption{We show average per-task agent run costs on the held-out set (judge cost is roughly constant across models and is excluded in the calculation) against micro-averaged task scores for each model. The Pareto frontier traces \textbf{Grok~4.20 $\to$ Gemini~3.1~Pro $\to$ GPT-5.5}; \textbf{Opus is fully dominated} -- GPT-5.5 achieves a similar overall score range at 5.6$\times$ lower cost ($\$8.85$ vs $\$49.21$ per task).}
\label{fig:pareto}
\end{figure}

Figure~\ref{fig:pareto} contextualizes task scores on the held-out set against agent run costs.

On the held-out set, the cost-quality frontier flattens at an average cost per task of around \$10: Gemini~3.1~Pro at \$5.78/task and GPT-5.5 at \$8.85/task land within 6.8~pp of Opus's macro score (53.5\%) but at 8.5-5.6$\times$ lower cost. The tradeoff between Gemini and GPT-5.5 hinges on rubric section: GPT-5.5 wins Formula Correctness (+9.9~pp), Model Integration (+6.4~pp), and Presentation (+24.4~pp); Gemini wins on Pitfalls (+18.1~pp, meaning fewer of its outputs trigger error-penalty criteria).

\subsection{Behavior Profiles}
\label{subsec:qualitative}

We also observe qualitatively distinct behavioral profiles of different language models. 

\paragraph{Opus is not uniformly cost-effective across task regimes.}
On a structurally large but routine balance-sheet construction task, Gemini 3.1 Pro scored 100\% at \$3.00, while Opus 4.7 scored 95\% at \$83.05. The trace suggests that Opus's inefficiency came from cost-dense interactions: broad workbook reads, large formula-writing calls, and substantially more output tokens. Gemini instead used many narrow \texttt{read\_range} probes, compact formula writes, programmatic formula copying, and recalculation checks to produce a better-scoring workbook at 27.7$\times$ lower cost. This suggests that high-cost reasoning models are not always preferable for well-specified workbook-wiring tasks.

\paragraph{GPT-5.5 tends to use \texttt{execute\_python} over structured tools.} On 64\% of manipulation tasks, GPT-5.5's first action after \texttt{get\_workbook\_state} was \texttt{execute\_python} to inspect or write multiple cells via openpyxl, vs $<$25\% for Opus, Sonnet, and Gemini. This explains GPT-5.5's lower turn count (24 vs 38 for Opus) but also its weaker performance on Output Validation criteria: code-driven writes require fewer turns but are harder to verify against the rubric's expected values without explicit recalculation tool calls. 

\paragraph{Sonnet trajectories include read-only actions on some tasks.} On $\sim$8\% of manipulation held-out tasks, Sonnet's output workbook was identical to the input (zero writes), despite a non-trivial trajectory of \texttt{get\_cells} and \texttt{read\_range} calls. The agent explored, decided to call \texttt{done}, and returned no modifications. We did not observe this on Opus, GPT-5.5, or Gemini at meaningful rates ($<$1\%).

\paragraph{Weaker models fail to call \texttt{recalc\_workbook} proactively.} Opus calls \texttt{recalc\_workbook} on 75\% of tasks, GPT-5.5 on 100\%, but Sonnet and Grok call it on $<$50\%, consistent with their weaker Perturbation pass rates. Models that do not recalculate workbooks cannot self-verify whether their dynamically-computed values are correct.

\subsubsection{Case Studies}
\label{subsec:case_studies}

We describe in Appx.~\ref{app:failure_anecdotes} a detailed walkthrough of three examples of tasks with different failure mode profiles across models, including side-by-side trajectory excerpts and error attribution to specific rubric criteria. The first case study describes a task that all models performed similarly poorly on. The second example describes a differentiating task where models scored along a gradient. Finally, the third example highlights a case with a 27.7x difference in cost between two similarly high-scoring models on a task.

\section{Limitations}
\label{sec:limitations}

Though tasks with multiple sheets per workbook are typical in our dataset, all tasks have a maximum of one spreadsheet workbook as context, with no additional artifacts as context (in contrast, other works have covered e.g. slide decks and retrieval of other documents as well).

We acknowledge the complex considerations around developing and evaluating AI system performance on end-to-end workflows with potential real-world utility. Though we discuss in \S\ref{sec:introduction} the potential for AI systems to aid finance professionals' productivity, truly scalable automation of major occupational responsibilities may have implications for job security and working conditions that are mixed or negative in effect.

The initial release of \benchname assumes a uniform harness designed to evaluate LLMs' underlying reasoning and instruction following capabilities over realistic task prompts that a finance professional might provide to an AI system, while providing enough scaffolding via tool use to allow for well-formed task attempts by most models evaluated. That being said, we acknowledge that many modern frontier models are specifically trained to perform well with a particular harness, and we plan in a future release to include an additional leaderboard that allow providers to submit endpoints assuming custom harnesses, including ones adapted from our open sourced harness. We invite readers to interpret our existing results on the benchmark as a strong baseline that invites iteration and improvement. 

\vspace{10pt}
\section{Conclusion}
\label{sec:conclusion}

We view our work as one piece of a larger effort to develop benchmarks for tasks aligned with real-world value; currently, representation in AI benchmarks is not in proportion with employment numbers \citep{wang2026ai4work}, with coding and mathematical tasks heavily overrepresented compared to other highly digital sectors. The existing benchmark representation skew is arguably unsurprising given the relative difficulty of collecting high quality, realistic examples for tasks reflective of real business value, and the relative challenge that scalable verification of task completion often presents. That being said, beyond our artifact contributions we hope that our methodological descriptions provide useful scaffolding for other researchers who may be interested in creating their own benchmarks or understanding the resources needed to do so.


\section*{Acknowledgements}

We thank Aidan Sheinberg, Filbert Presley, Daniel Fu, Ian Lee, Kasper Halevy, and Rishi Sethuraj for their contributions during the development of this work.




{
\small
\bibliographystyle{unsrtnat}
\bibliography{main}
}






\newpage
\appendix
\section{\benchnamelong Rubric Descriptions}
\label{app:bench_examples}
Synthesis and manipulation tasks in \benchname are evaluated with binary rubric criteria across six categories of items. We provide descriptions of our rubric criteria categories and examples of criteria in Table~\ref{tab:rubric_categories}.

\begin{table}[h]
\centering
\footnotesize
\begin{tabular}{L{1.4cm} L{3.5cm} L{7.8cm}}
\toprule
\textbf{Category} & \textbf{Description} & \textbf{Example criteria} \\
\midrule

Formula Correctness &
    A formula implements the intended financial calculation, crediting any methodologically valid formulation. &
    \textbf{(1 point)} Are the fiscal years (2021 through 2031) calculated by adding one to the prior year's column header (with only the first projected year hardcoded) on the Operating Model Cases tab? \newline
    \textbf{(2 points)} Is the LP ownership percentage for the first tier on the Waterfall tab calculated EITHER using both the LP equity contribution from the equity capital stack and the dilution from the promote OR as one minus the GP ownership percentage? \\
\addlinespace

Model Integration &
    A value is referenced from its authoritative source tab rather than recomputed or hardcoded locally. &
    \textbf{(5 points)} Is the WACC value in the assumptions block on the DCF tab linked to the WACC calculation on the WACC tab (not recalculated locally)? \newline
    \textbf{(2 points)} Is levered free cash flow on the Waterfall tab linked to the Full Monthly CF tab? \\
\addlinespace

Output Validation &
    A key output matches the reference answer within a stated tolerance. &
    \textbf{(3 points)} Is 2031 Revenue on the Operating Model tab within +/- 2\% of \$10,342 million? \newline
    \textbf{(5 points)} Is the Project Level IRR on the Waterfall tab within +/- 0.5 pp of 27.9\%? \\
\addlinespace

Perturbation & 
    Changing a driving input causes the dependent output to recompute to the correct new value. &
    \textbf{(5 points)} If the Active Case selector on the Operating Model Cases tab is changed from Base to Bull, does the Implied Share Price under the PGR Method on the DCF tab update to within +/- 2\% of \$255.11? \newline
    \textbf{(5 points)} If the GP Equity Contribution on the Waterfall tab is changed from 5.0\% to 10.0\%, does the GP IRR on the Waterfall tab update to within +/- 5pp of 56.1\%? \\
\addlinespace

Presentation &
    The model follows formatting and labeling conventions without altering underlying numeric values. &
    \textbf{(3 points)} Are the historical fiscal year columns labeled with an A suffix and the projected fiscal year columns labeled with a P suffix using number formatting (so the underlying values remain numeric)? \newline
    \textbf{(2 points)} Are hardcoded assumption values on the Waterfall tab displayed in blue font? \\
\addlinespace

Pitfalls & 
    Penalty checks for defects that degrade a model. & 
    \textbf{(-5 points)} Are there any error values across the Operating Model Cases, Operating Model, WACC, or DCF tabs (\#REF!, \#DIV/0!, \#VALUE!, \#NAME?, \#N/A, \#NUM!)? \newline
    \textbf{(-2 points)} Is any data or label text truncated due to insufficient column widths or row heights in the Waterfall tab? \\

\bottomrule
\end{tabular}
\caption{Rubric criteria categories with representative grading questions. Each criterion is a binary (yes/no) check worth a fixed point weight; weights vary with importance and difficulty. Pitfalls are scored as penalties (negative points); all other categories award positive points.} 
\label{tab:rubric_categories}
\end{table}

Note that we deliberately do not include a separate `structural completeness' section in our final rubric, though it was part of the initial design. Preliminary experimentation suggested that explicit structural‑completeness and input‑accuracy sections tend to over‑reward models for copying a particular template and over‑penalize valid alternative builds.
Rather than explicitly considering `structure' as its own rubric item category, we capture `structure' as a combination of formula correctness, model integration, output validation, and perturbation.

\section{Details of Contributor Recruitment, Onboarding, and Annotation Tasks}
\label{app:contributor_blurbs}

\paragraph{Outreach and eligibility.}
Admission to the platform was not open. Contributors joined through targeted outreach from our team or through referrals from existing contributors. Outreach was directed primarily at experienced finance professionals with one to twenty years of professional experience, with a smaller cohort of senior-year undergraduates who had completed prior investment banking or private equity internships. Contributors were drawn predominantly from top-tier U.S. universities, with additional representation from institutions in the United Kingdom, Ireland, India, Hong Kong, and continental Europe; in total, 78 contributors spanning seven countries participated in benchmark construction. Represented roles spanned investment banking (summer analyst through associate), private equity, hedge fund and asset management, management consulting, freelance and fractional finance, and senior positions including managing director, fractional CFO, and Chief AI Officer. Contributors were compensated at rates substantially above prevailing market rates for annotation work, reflecting the level of domain expertise required.

\paragraph{Application, screening, onboarding.} Because admission was gated at the outreach stage, contributors arriving at onboarding were largely pre-qualified for the work. Onboarding was structured to calibrate contributors to our task specification rather than to filter them, and combined asynchronous training material with a live calibration exercise with a member of our in-house finance team.

\begin{figure}[!ht]
        \centering
        \includegraphics[width=0.95\linewidth]{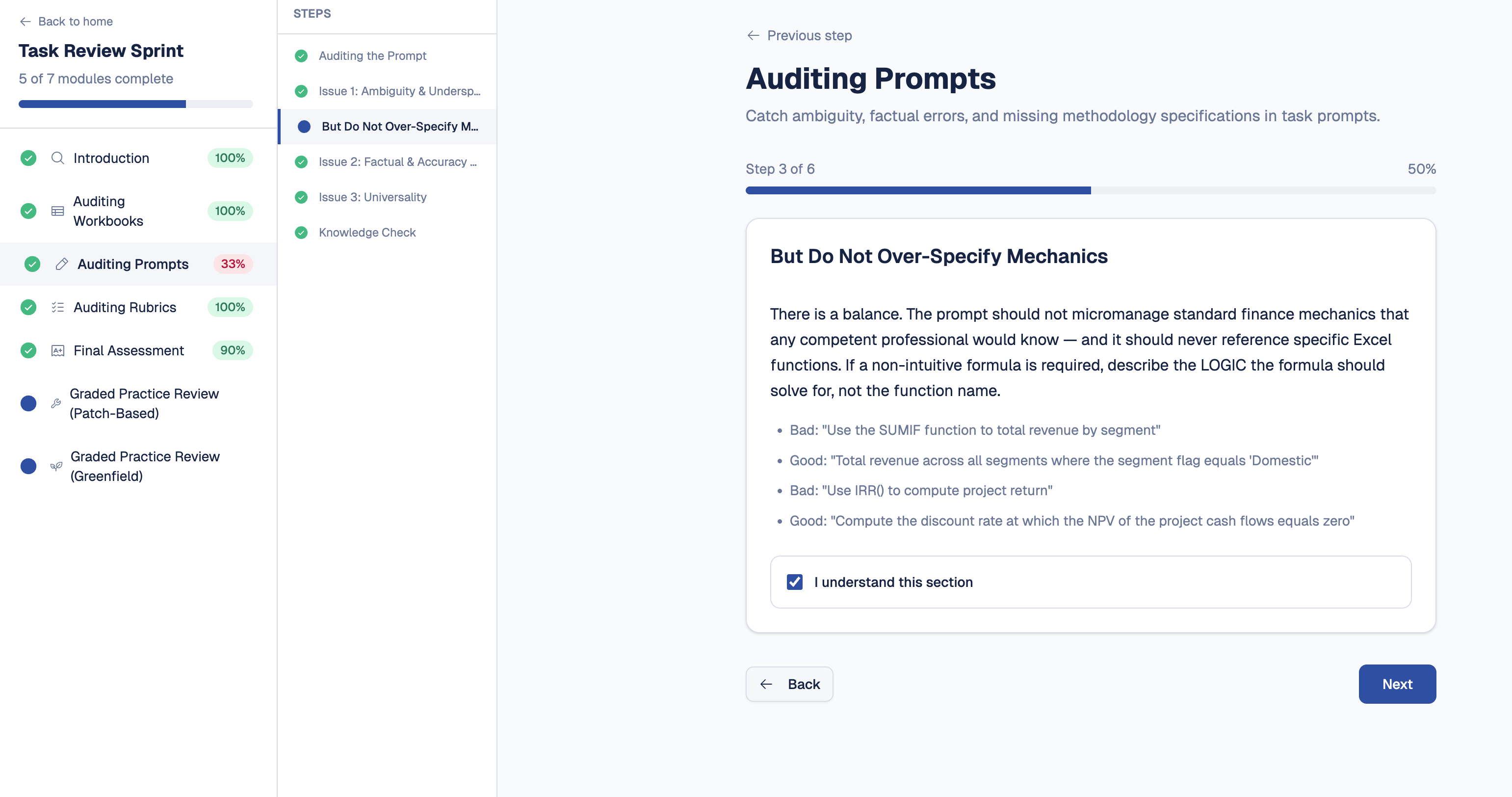}
        \caption{Sample module from the contributor onboarding sequence, which included training to calibrate understanding of common task-quality issues.}
        \label{fig:onboarding}
\end{figure}

\paragraph{Task construction and submission.}
After completing onboarding, contributors were authorized to create tasks. Submissions were required to pass automated quality checks before entering the review pipeline; these checks evaluated tasks across several quality-relevant dimensions to prevent malformed submissions from reaching reviewers.


\begin{figure}[h]
\centering
\begin{tcolorbox}[colback=gray!5!white, colframe=gray!50!black, boxrule=0.5pt, arc=2pt, width=0.95\linewidth]
... The AI will receive your starting spreadsheet and your natural-language prompt, attempt to produce the finished spreadsheet, and then be scored against your rubric. The better your tasks, the better the AI learns---your work directly shapes model intelligence on financial workflows.

\medskip

What you're making: an ``assignment'' for an AI spreadsheet analyst.
\begin{itemize}\setlength\itemsep{0.2em}
    \item \textbf{Input spreadsheet} = The ``before''---a starting workbook that needs work done to it.
    \item \textbf{Prompt} = The natural-language request---what a real user would type to ask an AI to transform the input into the output.
    \item \textbf{Output spreadsheet} = A strong example solution showing one correct way the task could be completed.
    \item \textbf{Grading Rubric} = A set of binary yes/no criteria that allows automated scoring of how well the AI's attempt matches the expected output.
\end{itemize}

Think of it like school: the prompt is the question, the output is a sample solution, and the rubric is the grading sheet. The model can get full credit even if its spreadsheet doesn't look identical, as long as it solves the problem and meets the rubric.

\medskip

\begin{tcolorbox}[colback=blue!5!white, colframe=blue!40!black, boxrule=0.5pt, arc=2pt]
\textbf{The Core Formula} \\
Input Spreadsheet \ + \ Prompt \ = \ Output Spreadsheet \ $\rightarrow$ \ Scored with Grading Rubric
\end{tcolorbox}

\textbf{Step-by-Step Instructions}
\begin{enumerate}\setlength\itemsep{0.2em}
    \item View task from the ``Submit Workbooks and Prompt'' category
    \item \ldots
\end{enumerate}
\end{tcolorbox}
\caption{Excerpt from the annotator instructions provided during onboarding. The full document covered task creation, rubric design, quality review, and calibration procedures.}
\label{fig:annotator-instructions-excerpt}
\end{figure}

\paragraph{Review. }
We held every submission to a rigorous quality bar, and the typical task underwent multiple revisions to reach acceptance. Reviewers were drawn from our in-house finance team or from top-performing contributors promoted into a reviewer role on the basis of demonstrated submission quality and command of the task specification. Reviewers worked through each submission against a structured checklist, returning the task with detailed feedback or, where edits were small, applying them directly. Only tasks that cleared this review were admitted into the final benchmark. 


\subsection{Annotation Burden Across the Task Life Cycle}
\label{app:annotation_hours}

\paragraph{Annotation burden and iterative quality refinement.} For the creation of tasks submitted to the benchmark, our team estimates an average of 5 hours per task, approximately evenly distributed across three components: initial authoring time, reviewer time, and the send-back and quality-control loop. 

The first captures contributor effort spent constructing the seed-to-output transformation, drafting the prompt, and assembling the supporting rubric criteria. The second captures reviewer time, working through each submission and engaging directly with the original contributor to bring less-experienced submissions up to specification. The third was a substantial additional portion of total annotation effort - driven not by the volume of tasks but by the iterative evolution of our internal quality standards during production. Over the course of the project, we innovated on our task quality along several dimensions, and each shift triggered retroactive rework on tasks that had already cleared earlier rounds of review. Some examples:

\begin{itemize}
    \item \textbf{Prompt design.} Early task prompts were overly prescriptive. We revised our prompt conventions to describe the desired output and the analytical intent, leaving the contributor (and, at evaluation time, the model) responsible for determining the path. A second-order effect of this change was that stripping away the procedural scaffolding exposed a complexity-calibration problem in the existing task pool.
    \item \textbf{Task complexity floor.} We tightened task complexity standards. Tasks falling below a certain threshold were either re-scoped or retired.
    \item \textbf{Rubric structure.} Early rubric drafts sometimes placed too much weight on surface-level indicators of task completion rather than on whether the submitted artifact satisfied the intended analytical requirements. We revised the rubric format to emphasize outcome correctness, robustness, and professional presentation, while preserving flexibility for multiple valid implementation approaches.
    \item \textbf{Workbook hygiene.} We added an automated pre-release layer of checks targeting environment artifacts (e.g., hidden sheets, stray named ranges).
\end{itemize}

A non-trivial share of previously accepted tasks were returned for rework following each of these shifts, in some cases more than once. 

\section{Human Utility Study}
\label{app:human_utility}

\begin{table}[h]
\centering
\begin{tabular}{lccc}
\hline
\textbf{Model} & \textbf{Reviewer 1} & \textbf{Reviewer 2} & \textbf{Total} \\
\hline
Total   & 55.56\% & 51.11\% & 53.33\% \\
\hline
Gemini  & 50.00\% & 62.50\% & 56.25\% \\
GPT 5.5 & 75.00\% & 62.50\% & 68.75\% \\
Grok    & 0.00\%  & 0.00\%  & 0.00\%  \\
Opus    & 75.00\% & 75.00\% & 75.00\% \\
Sonnet  & 62.50\% & 50.00\% & 56.25\% \\
\hline
\end{tabular}
\caption{``Yes'' proportion by reviewer and model.}
\label{tab:reviewer_utility}
\end{table}

Beyond rubric judgment, the usefulness of model outputs is itself worth measuring. To approach this task, we had two former finance professionals review outputs from our release set across both synthesis and manipulation tasks for Claude Opus 4.7, Claude Sonnet 4.6, GPT-5.5, Gemini 3.1 Pro Preview, and Grok 4.20.

For each model output, reviewers were asked to answer a single question: ``would this output substantially help in my build process, or would it require significant rework before I could continue?'' The intent is not to test whether models produce finished deliverables, but whether or not they add value to a professional's workflow. Structure, formulas, and formatting account for much of the manual effort in financial modeling, and even partial assistance in any of these areas saves real time. That being said, a model output that does not pass this check may reasonably be considered substantially flawed to the extent that a full retry or manual task completion might be warranted.

Results are reported in Table~\ref{tab:reviewer_utility}. Opus 4.7 was strongest on structure and formulas, and was the model reviewers most often flagged as a usable starting point for further work. Alternatively, Grok frequently failed to generate a complete output, which drags its scores down across both task types. Formatting was a consistent weak point across the board. In particular, models struggle to adhere to the formatting conventions of the workbook they are operating in, often defaulting to their own styling rather than matching what is already on the sheet.
\section{Public Set Scoring}
\label{app:public_set}
Table~\ref{tab:public_results} reports performance of an expanded set of models on only the public set.

\begin{table}[h]
  \centering
  \small
  \begin{tabular}{l|c|ccccc|cc}
  \toprule
  \textbf{Set} & \textbf{N} & \textbf{Opus} & \textbf{Sonnet} & \textbf{GPT-5.5} & \textbf{Gemini} & \textbf{Grok} & \textbf{Kimi} &  
  \textbf{MiniMax} \\
  \midrule
  Manipulation      & 7  & \textbf{59.4} & 48.6 & 57.1 & 45.9 & 10.4 & 45.0 & 6.0  \\
  Synthesis      & 1  & \textbf{77.0} & 45.0 & 45.0 & 44.0 & 34.0 & 0.0  & 38.0 \\
  Interrogation   & 3  & 50.0 & 50.0 & 50.0 & 50.0 & 50.0 & 50.0 & 50.0 \\
  \midrule
  \textbf{Overall (micro)} & \textbf{11} & \textbf{58.4} & 48.7 & 54.1 & 46.8 & 23.3 & 42.3 & 20.9 \\
  \bottomrule
  \end{tabular}
  \caption{Public set performance (\%, micro-aggregate weighted by task count). Frontier models (Opus, Sonnet, GPT-5.5, Gemini, Grok)   
  and OSS models (Kimi K2.6, MiniMax M2.5) on 11 public tasks released for reproducibility.}
  \label{tab:public_results}
  \end{table}

\section{Harness Implementation Details}
\label{app:harness_methodology}

In \S\ref{sec:evaluation} we focus on motivating the design decisions underlying our open source harness. We detail our system design and implementation below.

\subsection{System instructions}
\label{app:si_harness}

We share our harness's minimal SI, which, like the harness itself, we designed to evaluate underlying reasoning abilities without a bias towards any specific approach that different models might take. 
\begin{lstlisting}[breaklines=true,basicstyle=\ttfamily\footnotesize,frame=single,xleftmargin=0pt]
You are an expert finance professional. You will be given a task and a set of tools to work with a spreadsheet. When finished, call 'done'.
All tool responses return JSON with a 'status' field ('ok' or 'error').
Use Excel formulas to keep the spreadsheet dynamic and auditable. Do not hardcode computed values when a formula can produce the result.
Large tool results may be truncated. Read specific ranges rather than entire sheets.
\end{lstlisting}


On the other hand, our evaluation agent's system prompt is more prescriptive -- however, we evaluate only the final output rather than also considering detailed agent trajectories.

\begin{lstlisting}[breaklines=true,basicstyle=\ttfamily\small,frame=single,xleftmargin=0pt,columns=fullflexible]
You are an expert spreadsheet evaluation judge. You will be given an Excel workbook and a list of evaluation criteria. Your job is to carefully inspect the workbook -- its sheets, cell values, formulas, structure, formatting, and organization -- and determine whether each criterion is met.

Guidelines:
- Open and thoroughly inspect the workbook before answering.
- For criteria about specific values, check the actual cell values (use data_only mode mentally -- report the computed value, not the formula text).
- For criteria about formulas/references, inspect whether cells use formulas that reference the expected sources (vs hardcoded values).
- For criteria about structure (tabs, sections, line items), check sheet names and scan for the described layout.
- For criteria about dynamic behavior ("when X changes, does Y update"), check whether Y's formula references X either directly or transitively.
- For criteria about formatting or presentation (number formats, bold/italic, colors, borders, conditional formatting, data validation, column widths, hidden rows/columns, merged cells, charts, or visual layout), actively inspect the relevant cell or range formatting properties. Do not assume formatting is correct without checking.
- Be precise. If a criterion asks for a value within a tolerance, compute the actual deviation.
- Every criterion is binary: met or not met. If the referenced sheet or cell does not exist, the criterion is NOT MET. If you cannot find the expected structure, the criterion is NOT MET. There is no middle ground -- make a judgment call.

PITFALL AND ERROR DETECTION:
- For penalty criteria (negative-weight questions about errors, hardcoded values, or broken references), you MUST actively search for the described problem rather than assuming the workbook is clean.
- Specifically check for: spreadsheet errors (#REF!, #DIV/0!, #VALUE!, #NAME?, #N/A), hardcoded numeric values where formulas are expected, broken or missing cross-sheet references, and circular reference errors that did not converge.
- When checking whether a cell contains a formula or a hardcoded value, always use read_range to inspect BOTH the computed value AND the formula string. A cell displaying $100,000 could be the hardcoded number 100000 (bad) or the formula =SUM(B2:B5) (good). You must distinguish these.
- When a criterion is a penalty question phrased as "Does the workbook contain errors/external links/hardcoded values?", answer true if the condition IS present (i.e., the penalty applies), false if it is NOT present (i.e., clean workbook).
- Scan broadly for errors: do not limit your inspection to only the cells mentioned in the rubric. If a criterion asks about errors in a section, check all cells in that section.

PERTURBATION TESTING:
- For criteria that test dynamic behavior ("if input X changes, does output Y update correctly"), perform the following steps:
  1. Read the current value of the target output cell.
  2. Modify the specified input cell to the test value.
  3. Call recalc_workbook to recompute all formulas.
  4. Read the target output cell again and verify it changed as expected.
  5. Restore the original input value and recalc to leave the workbook unchanged.
- If the output does not change after the input modification, the criterion is NOT MET (the model likely hardcoded the value).

For each criterion, respond with:
  - criterion_id: the exact ID string provided
  - met: true if the criterion's condition is satisfied, false otherwise
  - evidence: a brief (1-3 sentence) explanation citing specific sheets, cells, or values you observed. NEVER leave evidence empty -- always explain what you checked and what you found.

CRITICAL OUTPUT INSTRUCTION: When you have finished inspecting the workbook and evaluated all criteria, you MUST call the "done" tool and pass your evaluations as the "answer" parameter. The answer MUST be a JSON object in this exact format:

{"evaluations":[{"criterion_id":"...","met":true,"evidence":"..."},{"criterion_id":"...","met":false,"evidence":"Sheet 'DCF' not found in workbook"},...]}

Do NOT call done() without an answer. Do NOT put the evaluations in your message text. The ONLY way to submit your evaluations is done(answer='{"evaluations":[...]}').
\end{lstlisting} 


Interrogation tasks are evaluated with a different SI:

\begin{lstlisting}[breaklines=true,basicstyle=\ttfamily\footnotesize,frame=single,xleftmargin=0pt,columns=fullflexible]
You are grading a financial modeling answer. Your job is to compare the model's answer to the expected answer at the component level.

QUESTION:
{question}

EXPECTED ANSWER:
{expected_answer}

MODEL'S ANSWER:
{model_answer}

STEP 1 - Identify components.
Break the expected answer into its distinct value components. Each independent value, quantity, date, or yes/no judgment is a separate component. Ignore parenthetical annotations, source notes, or commentary in the expected answer - only extract the actual answer values.

Examples:
  "83 Months (Nov-27) and $35,804,564" -> two components: "83 Months (Nov-27)" and "$35,804,564"
  "$2.4mm" -> one component: "$2.4mm"
  "No, taxable losses would exceed taxable income" -> two components: "No" and "taxable losses would exceed taxable income"
  "274.8 (note difference in cash flow..." -> one component: "274.8" (the parenthetical is an annotation, not part of the answer)

STEP 2 - Grade each component.
For each component, determine whether the model's answer contains a matching value. Apply these rules:

  FORMAT NORMALIZATION (allowed):
  - $35,804,564 = 35804564 = 35.8M = $35.8 million
  - 25.4% = 0.254
  - 83 Months = 83 months
  - 2.4mm = $2,400,000 = 2.4 million
  - Week 12 = week 12

  NUMERIC TOLERANCE:
  - Values must match to within 1% relative difference OR differ only in trailing decimal precision. Dates must be exact matches.
  - 13.79 matches 13.78. 42.1705 matches 42.17.
  - 108,061 does NOT match 965,000.
  - If the model gives extra decimal precision beyond the expected answer, that is fine (0.0758 matches 0.07581570869).

  TEXT / QUALITATIVE:
  - Yes/No and directional answers (increase/decrease) must match exactly in meaning, including the positive / negative meaning of a number.
  - Minor phrasing differences are OK if the meaning is identical.

  EXTRACTION:
  - The model's answer may contain extra text, explanation, or working. Extract the relevant value and compare.
  - If the model hedges ("approximately $35M") but the value is within tolerance, it matches.
  - If the model explicitly states it cannot compute the answer, that component does not match.

STEP 3 - Return your assessment as a JSON object. Nothing else.

{"components": [
    {"expected": "<component 1>", "match": true/false, "reasoning": "brief explanation"},
    {"expected": "<component 2>", "match": true/false, "reasoning": "brief explanation"}
  ],
  "score": <number of matching components / total components>,
  "reasoning": "one-sentence overall summary"
}
\end{lstlisting}

\subsection{Tool catalog}
\label{app:tools}

\begin{itemize}
  \item \emph{Read.} \texttt{get\_cells} (sheet preview, configurable
        row count); \texttt{read\_range} (specific rectangular range);
        \texttt{get\_sheets} (sheet list with used ranges).
  \item \emph{Write.} \texttt{set\_cells} (batch values or formulas;
        per-cell or per-range mode); \texttt{create\_sheet},
        \texttt{delete\_sheet}; \texttt{insert\_rows},
        \texttt{insert\_columns}, \texttt{delete\_rows},
        \texttt{delete\_columns}.
  \item \emph{Format.} \texttt{set\_cell\_format} (bold, italic, font
        size/color, background, number format, alignment, border;
        single-range or batch); \texttt{merge\_cells},
        \texttt{unmerge\_cells}; \texttt{set\_column\_width},
        \texttt{set\_row\_height}.
  \item \emph{Other.} \texttt{auto\_filter}; \texttt{create\_chart}
        (bar / line / pie / scatter); \texttt{execute\_python}.
  \item \emph{Virtual}:
        \texttt{get\_workbook\_state} (sheet names, active sheet,
        per-sheet dimensions); \texttt{done} (termination signal).
\end{itemize}

Tool schemas are defined once in JSON Schema in a single registry (\texttt{TOOL\_REGISTRY}). All adapters convert from this canonical form; none re-author tool semantics.

\subsection{\texttt{execute\_python} sandbox}
\label{app:sandbox}

Three layers of defense:
\begin{enumerate}
  \item \emph{Builtin allow-list.} A curated set of safe builtins (types,
        iteration, math, \texttt{print}, common exceptions) is exposed.
        \texttt{open}, \texttt{exec}, \texttt{eval}, \texttt{compile},
        \texttt{input}, \texttt{breakpoint}, \texttt{exit}, \texttt{quit},
        \texttt{help} are replaced with stubs that raise
        \texttt{PermissionError}.
  \item \emph{Import allow-list.} A wrapper around \texttt{\_\_import\_\_}
        permits only \texttt{math}, \texttt{datetime}, \texttt{decimal},
        \texttt{fractions}, \texttt{statistics}, \texttt{collections},
        \texttt{itertools}, \texttt{functools}, \texttt{string},
        \texttt{re}, \texttt{copy}, \texttt{json}, and \texttt{openpyxl.*}
        submodules. Imports of \texttt{os}, \texttt{subprocess},
        \texttt{sys}, \texttt{socket}, etc., raise \texttt{ImportError}.
  \item \emph{Wall-clock timeout.} A 30-second \texttt{SIGALRM} kills
        runaway code.
\end{enumerate}

\subsection{Provider adapters}
\label{app:adapters}

One adapter per provider, each implementing the same
\texttt{act(observation, tool\_schemas) $\to$ ToolCall} interface: 
Anthropic Messages, OpenAI Chat Completions, OpenAI Responses (the newer reasoning-model API), Google Gemini, and OpenAI-compatible endpoints (used for Fireworks-hosted GLM-5, MiniMax, Kimi). A vLLM stub is present for future use.

Provider-specific notes that affect comparability:
\begin{itemize}
  \item \emph{Anthropic adaptive thinking.} When \texttt{reasoning\_effort} is set, the API requires \texttt{temperature=1}; the adapter silently overrides any configured temperature in this case.
  \item \emph{OpenAI Responses reasoning summaries.} Capturing reasoning output requires \texttt{summary: "detailed"}; without it, the \texttt{thinking} field is always \texttt{None}.
  \item \emph{Multiple tool calls.} Anthropic models occasionally emit multiple \texttt{tool\_use} blocks per response. The adapter keeps the first and strips the rest from history; the API would otherwise reject subsequent requests for missing \texttt{tool\_result} matches. The OpenAI Responses adapter applies the analogous filter to \texttt{function\_call} items.
  \item \emph{Reasoning-effort scales are not cross-provider.} Anthropic's \texttt{\{low, medium, high, max\}} and OpenAI's \texttt{\{low, medium, high\}} share names but are not calibrated to each other. Results are reported per-model.
  \item \emph{Rate-limit handling.} Up to 8 retries with exponential backoff and best-effort parsing of provider-supplied retry hints.
\end{itemize}

\subsection{Termination mechanism}

If a model response contains no tool call, the adapter appends a single
re-prompt (``Please take an action by calling one of the available tools,
or call \texttt{done} if you have finished the task'') and re-issues the
request. If the second response also lacks a tool call, the adapter
returns \texttt{ToolCall(name="done", arguments=\{\})} and the run
terminates.

\subsection{Observation truncation}

Tool results are capped at 30{,}000 characters. Strategy:
\begin{enumerate}
  \item If under cap, untouched.
  \item Otherwise, identify large array fields (\texttt{values}, \texttt{formulas}, \texttt{preview}); keep first 40\% and last 20\% of rows with an elision marker for the omitted middle. 
  \item If still oversized, hard-truncate the JSON string with a marker.
\end{enumerate}
Truncation sets \texttt{\_truncated: true} on the observation. The character cap follows the precedent set by \texttt{mini-swe-agent}'s 10K cap, relaxed for spreadsheet-shaped results.

\subsection{Trajectory log format}

One JSON-Lines file per run (\texttt{\{task\_id\}\_\{model\}\_\{timestamp\}.jsonl}), flushed line-by-line. Entry types:
\texttt{system} (turn 0; task prompt, prompt version and hash),
\texttt{action} (per turn; tool, args, agent thinking, API latency, per-turn input/output tokens, per-turn and cumulative cost), 
\texttt{observation} (per turn; tool result, execution time),
\texttt{summary} (final totals). Schema is a single Pydantic model with all fields optional and \texttt{exclude\_none=True} on serialization; field additions are backwards compatible.

\subsection{Cloud execution}

The Modal-based runner schedules one serverless container per
(\texttt{task}, \texttt{model}) pair. Provider-pinned worker pools (\texttt{anthropic}, \texttt{openai}, \texttt{gemini}) carry only their own API key secret and have independent \texttt{max\_containers} caps, so different providers' rate limits cannot interfere. Submission is asynchronous fan-out (\texttt{spawn} for every job, then collect as results complete); per-task wall-clock timeout is 30 minutes with two retries. Results land in a Modal Volume packaged as \texttt{\{study\}/\{model\}\_\{task\_id\}/} with \texttt{input/} and \texttt{output/} subdirectories. Failures are persisted (partial trajectory, error type, traceback) rather than raised, so a batch is fault-tolerant.

\subsection{Replay and inspection}

Trajectories are sufficient to replay every workbook mutation offline. A standalone visualizer re-executes each tool call against a fresh \texttt{openpyxl} instance, snapshots state after every mutating call, and emits a self-contained HTML viewer (no JavaScript dependencies, no network) with a tool-call timeline, before/after diff highlighting, and captured reasoning text per turn. This decouples post-hoc inspection from the live API and lets reviewers audit a run without re-incurring inference cost.

\section{Benchmarking Costs}
\label{app:cost_tokens}
Table~\ref{tab:cost_tokens} describes usage and cost metrics associated with model evaluations on \benchnamelong. The average cost per task instance attempt is roughly \$24, and considering evaluation costs increases the per-task cost to nearly \$34 -- reflecting the medium to long horizon task instances in the benchmark.

\begin{table}[h]
\centering
\small
\setlength{\tabcolsep}{5pt}
\begin{tabular}{l|rr|r|rr}
\toprule
\textbf{Model} & \textbf{Agent} & \textbf{Judge} & \textbf{Total} & \textbf{Input} & \textbf{Output} \\
& \textbf{Cost} & \textbf{Cost} & \textbf{Cost} & \textbf{Tokens (M)} & \textbf{Tokens (M)} \\
\midrule
Opus 4.7    & \$8{,}070 & \$1{,}116 & \$9{,}186 & 513.2 & 4.96 \\
Sonnet 4.6  & \$3{,}469 & \$1{,}189 & \$4{,}658 & 1{,}127.6 & 5.76 \\
GPT-5.5 & \$1{,}451 & \$1{,}292    & \$2{,}743   & 274.1 & 2.69 \\
Gemini 3.1 Pro  & \$914 & \$1{,}072 & \$1{,}986 & 717.0 & 1.76 \\
Grok 4.20   & \$614 & \$1{,}071    & \$1{,}685  & 304.7 & 0.78 \\
\midrule
\textbf{Total}  & \textbf{\$14{,}518} & \textbf{\$5{,}740} & \textbf{\$20{,}258} & \textbf{2{,}936.6} & \textbf{15.95} \\
\bottomrule
\end{tabular}
\caption{Total cost and token usage to evaluate all five frontier models on the \benchname held-out set (120 tasks: 75 manipulation, 9 synthesis, 36 interrogation). Agent cost refers to the model under evaluation; judge costs are from the GPT-5.4 grader.
Token counts reflect agent input/output across all task runs; Sonnet's high input volume reflects shorter context but more turns per task, while Opus's high output volume reflects extended thinking.}
\label{tab:cost_tokens}
\end{table}
\section{Failure Modes Case Studies}
In this section, we detail three case studies with varying comparative performance profiles in different models.

\label{app:failure_anecdotes}
\subsection{Case Study 1: Difficult task where all models fail to achieve satisfactory performance}
We describe a task where every endpoint failed to generate useful spreadsheets (maximum score: 35\%), with the failure mode bifurcating cleanly between the two dominant primary buckets (\texttt{B5 Workbook Integration Failure} and \texttt{B7 Conceptual Financial Logic Error}). This is a canonical ``shared challenge with diverging mechanism'' pattern.

\paragraph{Task Overview.} The model was provided with an existing financial model containing populated \texttt{Inputs}, \texttt{Capitalization}, and \texttt{PPR} (Production / Reserves) tabs and tasked with constructing a new \texttt{Merger Model} tab covering 2021E and 2022E. The structure required four column groups (\texttt{EQT Status Quo}, \texttt{CHK Status Quo}, \texttt{Merger Adjustments}, \texttt{EQT Pro Forma}) populated with EBITDA, Interest Expense on Existing Debt, Interest Expense on New Debt, Taxes, and Other to produce Cash Flow from Operations; a Cash Flow section bridging CFO to Free Cash Flow; a Capitalization section computing total debt; Credit Statistics (Debt/EBITDA, Net Debt/EBITDA, Debt/Production); a Liquidity section bridging Revolver Size to Total Liquidity; a revolver build section reconciling Beginning Balance through Borrowings, Repayments, and Ending Balance; and a Cash Schedule reconciling beginning cash through Use of Cash and Accretion to ending cash. The critical structural requirement -- the one that drove this task's universal failure -- was that line items shared with the existing model (Shares Outstanding, Senior Notes, Corp Adj., Interest Expense on Notes, Letters of Credit) be \emph{linked} to the \texttt{Capitalization} and \texttt{Inputs} tabs via formulas that aggregate conditionally on the existing schedules' roll/flag columns, rather than recalculated locally on the \texttt{Merger Model} tab from raw rate assumptions.

\paragraph{Sample 1: Gemini 3.1 Pro (Score: 22.2\%).}Top failed criterion: \texttt{FAILS\_UNDER\_INPUT\_CHANGE}. EQT Status Quo EBITDA linked directly to hardcoded input values (\texttt{K44}) in the \texttt{Inputs} tab. While this yielded the correct answer under the current acquirer configuration, the linkage failed to generalize dynamically if the acquirer changed. The model therefore required manual formula edits to maintain correctness. The intended implementation was to link to columns E and F on the \texttt{Inputs} tab, which dynamically pull formulas based on the selected acquirer. The model thus failed not on arithmetic correctness but on integration robustness and workbook dynamism.

\paragraph{Sample 2: Claude Opus 4.7 (Score: 30\%).} Top failed criteria:

\begin{itemize}
    \item \texttt{WRONG\_DEBT\_ROLLFORWARD}. Revolver borrowings and repayments were driven off \texttt{Change in Cash} (\texttt{C32}) using:
    \begin{verbatim}'Merger Model'!C56 = MAX(0,-C32)'Merger Model'!C57 = -MAX(0,C32)\end{verbatim}
    This formulation is partially circular because \texttt{C32} already includes revolver activity. Borrowings and repayments should instead have been driven from free cash flow availability. The resulting revolver schedule therefore produced mechanically incorrect debt flows.    
    \item \texttt{WRONG\_CELL\_REFERENCE}. Cash Flow from Investing referenced row 47 of the \texttt{Inputs} tab, which corresponds to 2021 Daily Production Volume rather than Capex. The correct implementation required linking to the Capex row highlighted in the workbook.
\end{itemize}

\subsection{Case Study 2 (Differing performance across models)} 

We describe a task example where one defect class -- off-by-one row references against the \texttt{Inputs} tab -- sorts five models into a clean four-tier outcome gradient, illustrating how referential discipline drives most differential scoring.

\paragraph{Task Overview.} The model was provided with an existing operating model containing an \texttt{Inputs} tab (with annual operating-expense line items in column D), a KPI Dashboard, and an Income Statement, and tasked with constructing a new \texttt{Operating Expenses} tab. The task required:

\begin{enumerate}
    \item Building a monthly OpEx schedule from January 2024 through December 2028 by dividing each annual \texttt{Inputs} value by 12;
    \item Computing annual subtotals as the sum of the 12 monthly cells;
    \item Wiring the new tab back into the KPI Dashboard's ``other opex'' placeholders and the Income Statement operating-expense block; and
    \item Preserving dynamic flexibility under perturbation: if \texttt{Office Lease Rental} (\texttt{Inputs} row 68) changes, the corresponding Income Statement period must update without manual intervention.
\end{enumerate}

The task is structurally simple -- no novel financial logic is required -- but stresses a specific capability: locating the correct row in a moderately deep input schedule and propagating it consistently through approximately 200 calculated cells across a five-year monthly horizon.

\paragraph{Sample 1: Gemini 3.1 Pro (Score: 32\%).}Top failed criteria:

\begin{itemize}
    \item \texttt{WRONG\_CELL\_REFERENCE}
    \item \texttt{FAILS\_UNDER\_INPUT\_CHANGE}
\end{itemize}

The \texttt{Office Lease Rental} row on the \texttt{Operating Expenses} tab linked to a blank cell on the \texttt{Inputs} tab rather than the correct line item, causing the value to resolve to zero. Every subsequent OpEx line item was then sourced from the wrong \texttt{Inputs} row by an offset of one row. Additionally, the formulas were not implemented dynamically with date-aware references, producing further integration fragility. This propagated into a downstream-output failure where both the Income Statement and KPI Dashboard inherited incorrect values from the malformed OpEx schedule.

\paragraph{Sample 2: Claude Opus 4.7 (Score: 64\%).} Top failed criteria:

\begin{itemize}
    \item \texttt{WRONG\_CELL\_REFERENCE}
    \item \texttt{FAILS\_UNDER\_INPUT\_CHANGE}
    \item \texttt{DOWNSTREAM\_OUTPUT\_FAILURE}
\end{itemize}

Like Gemini, Opus linked OpEx line items to incorrect rows in the \texttt{Inputs} tab and implemented them using direct cell links rather than dynamically aggregated formulas.An additional failure category was:
\begin{itemize}
    \item \texttt{PARTIAL\_IMPLEMENTATION}. The prompt required the Income Statement operating-expense block to source values from the newly created \texttt{Operating Expenses} tab. Eleven of thirteen lines did so correctly:
    \begin{verbatim}'Income Statement'!H17:H27 ='Operating Expenses'!H10:H20\end{verbatim}
    However, two lines bypassed the new schedule entirely:
    \begin{verbatim}
    'Income Statement'!H16 = Payroll!J101
    'Income Statement'!H28 = 'Depreciation Schedule'!D40\end{verbatim}
    Salaries therefore linked directly to Payroll, and Depreciation linked directly to the Depreciation Schedule, rather than routing through the new \texttt{Operating Expenses} tab as instructed.
\end{itemize}

Even if the row references had been correct, the implementation still relied on direct hard-linked formulas rather than year-aware dynamic aggregation logic, reducing model flexibility and maintainability.

\paragraph{Sample 3: GPT-5.5 (Score: 88\%).} This was the cleanest of the three OpEx-tab implementations because it correctly mapped the \texttt{Inputs} rows -- the single failure that both Gemini and Opus reproduced. The model maintained proper propagation into downstream statements and dashboards and preserved workbook dynamism under perturbation testing.

\subsection{Case Study 3: Same score, different costs}
We highlight the highest cost-ratio pair, where two endpoints produced functionally equivalent balance sheets at a 27.7$\times$ cost differential. This example is evidence that average-quality scoreboards can substantially understate divergence in cost-per-correct-output across endpoints.

\paragraph{Task Overview.} The model was provided with an existing operating model for Apex Technologies Ltd. and tasked with building a new \texttt{Balance Sheet} tab covering historical years F2017A--F2022A and forecast years F2023E--F2030E (14 years total). The task specified an exact line-item structure:

\begin{itemize}
    \item A Non-Current Assets section (12 rows including PP\&E, intangibles, deferred tax, and three Trade Receivables splits);
    \item A Current Assets section (10 rows);
    \item An Equity \& Liabilities section with Equity, Non-Current Liabilities, and Current Liabilities subsections;
    \item Top-level subtotals rolling into Total Equity and Liabilities; and
    \item A Check line equal to Total Assets minus Total Equity and Liabilities, which must equal zero across all 14 years.
\end{itemize}

All values had to be dynamically linked from supporting schedules within the workbook: no hardcodes, full workbook integration, and exact balancing within tolerance.

\paragraph{Cost and Scores.} 
\begin{center}
\begin{tabular}{lcc}
\toprule
Model & Cost & Score \\
\midrule
Gemini 3.1 Pro & \$3.00 & 100\% \\
Claude Opus 4.7 & \$83.05 & 95\% \\
\bottomrule
\end{tabular}
\end{center}

This produces a cost ratio of 27.7$\times$. Both models generated structurally correct Balance Sheet tabs satisfying the core prompt requirements: historical F2017A-F2022A and forecast F2023E-F2030E columns, the full requested asset taxonomy (PP\&E, CWIP, ROU, Goodwill, intangibles, investments, billed/unbilled receivables, loans, other financial assets, income tax, deferred tax, and other assets), the requested Equity and Liabilities ordering, the three top-level subtotals, Total Equity and Liabilities, and a Check line. The cost-performance tradeoff strongly favors Gemini. Moreover, the Opus-generated balance sheet did not fully balance and contained one truncated column. This task is structurally large but algorithmically routine: it requires exhaustive, accurate workbook wiring rather than novel financial reasoning. That regime appears well matched to Gemini-class models. The evidence therefore suggests reserving high-cost reasoning models such as Opus for tasks where incremental quality reflects genuine reasoning depth -- ambiguous prompts, novel financial structures, or multi-step debugging -- rather than for high-specification pattern execution.



\end{document}